\documentclass[
aps,
prd,
10pt,
twocolumn,
nofootinbib,
preprintnumbers,
superscriptaddress,
showpacs]{revtex4-2}

\usepackage{comment}
\usepackage{booktabs,makecell}
\usepackage{graphicx}
\usepackage{amsmath,amssymb}
\usepackage{amsfonts}
\usepackage{xspace} 
\usepackage[usenames]{color}
\usepackage{dcolumn}
\usepackage{bm}
\usepackage{mathrsfs}
\usepackage[colorlinks=true,citecolor=blue]{hyperref}
\usepackage[all]{hypcap} 
\usepackage[utf8]{inputenc} 
\usepackage{slashed}
\usepackage{multirow}
\usepackage{rotating}
\definecolor{orange}{rgb}{1,0.5,0}
\usepackage{tabularx}
\usepackage{siunitx}
\usepackage[normalem]{ulem}
\usepackage[dvipsnames]{xcolor}

\def\Mb{M_{\rm b}}
\def\epsc{\varepsilon_{\rm c}}

\begin{document}

\title{Phase transition structure of scalarized neutron stars: the effect of rotation and linear coupling}

\author{Kalin V. Staykov}\email{kstaykov@phys.uni-sofia.bg}
\affiliation{Department of Theoretical Physics, Faculty of Physics,
Sofia University ``St. Kliment Ohridski", Sofia 1164, Bulgaria}

\author{Fethi M. Ramazano\u{g}lu}
\email{framazanoglu@ku.edu.tr}
\affiliation{Department of Physics, Ko\c{c} University, Rumelifeneri Yolu,
34450 Sar{\i}yer, \.{I}stanbul, T\"{u}rkiye}

\author{Daniela D. Doneva}\email{daniela.doneva@uni-tuebingen.de}
\affiliation{Departamento de Astronom\'ia y Astrof\'isica, Universitat de Val\`encia,
Av. Vicent Andrés Estellés, 19, 46100, Burjassot (Val\`encia), Spain}
\affiliation{Theoretical Astrophysics, Eberhard Karls University of T\"ubingen, 72076 T\"ubingen, Germany}

\author{Stoytcho S. Yazadjiev}\email{yazad@phys.uni-sofia.bg}
\affiliation{Department of Theoretical Physics, Sofia University ``St. Kliment Ohridski", Sofia 1164, Bulgaria}
\affiliation{Institute of Mathematics and Informatics, Bulgarian Academy of Sciences, Acad. G. Bonchev St. 8, Sofia 1113, Bulgaria}

\begin{abstract}
There has been a recent revival in understanding the spontaneous scalarization phenomenon in scalar-tensor gravity as a phase transition. Using the tools of the Landau theory, we now know that first-order transitions where scalarization occurs in a discontinuous manner is more prominent than what had been considered in the literature, and this might lead to novel observation channels. However, the examples so far have been restricted to specific quadratic scalar coupling terms and spherically symmetric stars. Here we explore the phase transition structure of scalarization for more general couplings, considering linear as well as quadratic terms in the conformal scaling factor of the theory. Moreover, we also investigate the effect of rotation on the scalarization phase transition. Both of these considerations are natural choices since the coupling in a scalar-tensor theory can appear at all orders, and astrophysical neutron stars commonly have angular momentum. The introduction of linear coupling leads to a complex solution space which is harder to explore. However, we demonstrate that the Landau model of scalarization enables us to systematically find the branches of scalarized solutions that are commonly overlooked in numerical searches, providing a novel tool. On the other hand, the main effect of stellar rotation is shifting the stellar masses at which the phase transition occurs to higher values, but the qualitative picture remains similar to what happens under spherical symmetry.
\end{abstract}

\maketitle

\section{Introduction}

Spontaneous scalarization is a phenomenon in scalar-tensor type of generalizations of Einstein's gravity where many astrophysical systems behave very closely to what would happen in general relativity (GR), but this is not necessarily the case for some strongly gravitating systems~\cite{Damour:1992we, Damour:1993hw}. A tachyonic instability is incited once certain conditions are satisfied, leading to a highly \emph{scalarized object} which is typically stable and energetically favorable over its bald, zero-scalar-field counterpart. Such a process can produce large deviations from GR, which is one of the main reasons it has been intensely studied during the last few decades \cite{Doneva:2022ewd}. Going beyond the original work of Damour and Esposito-Far{\`e}se (DEF) where only neutron star scalarization was considered~\cite{Damour:1993hw}, now we know that all kinds of compact objects, including black holes~\cite{Doneva:2017bvd,Silva:2017uqg,Antoniou:2017acq} and even boson stars \cite{Whinnett:1999sc,Alcubierre:2010ea,Brihaye:2019puo}, can develop scalar field clouds in different \emph{models} of scalarization.

Scalarization can be investigated as a phase transition, which was realized in its early days, but this view usually took a secondary place in understanding the phenomenon~\cite{Damour:1996ke}. Moreover, it has been almost exclusively understood to be a second-order phase transition, where the move from an unscalarized (GR) solution to a scalarized one appears in a continuous manner, in the sense that one can, in principle, observe stable objects with arbitrarily weak scalarization and arbitrarily close structure to those in GR. There have been exceptions where discontinuous first-order scalarization has been observed, but mainly in quite specific scenarios~\cite{Kuan:2022oxs,Doneva:2021tvn,Doneva:2022yqu,Doneva:2023kkz,Pombo:2023lxg, Staykov:2024jbq}.\footnote{Among these references, we want to highlight \textcite{Kuan:2022oxs} as a not-so-specific scenario. Scalarized solutions typically transition to GR in two different places, at the low mass where scalarization commences and at the high mass where it ceases to exist. We will mainly study the low mass region in this work, whereas \textcite{Kuan:2022oxs} investigated the high mass.} 

The phase transition picture of scalarization recently changed when \textcite{Unluturk:2025zie} showed that first-order (discontinuous) phase transitions are quite common even in the original, and arguably the simplest, model of scalarization by DEF. All one needs is to consider slightly different coupling constants. Even more surprisingly, the DEF model and its extensions to massive scalars~\cite{Ramazanoglu:2016kul,Yazadjiev:2016pcb} are now known to feature first-order scalarization more commonly than the second-order version in the sense that first-order scalarization occurs in a larger region of the parameter space. This is exciting because it allows the coexistence of locally stable scalarized and unscalarized configurations at the same baryon mass, and opens up the possibility of transitions between them when there are sufficiently large astrophysical perturbations. This was quickly followed by other work that investigated unexpected phenomena like negative susceptibility~\cite{Muniz:2025egq} and phase transition structure of more exotic objects~\cite{Huang:2025dgc}. The first- and second-order phase transition structure of black hole scalarization was very recently investigated in \cite{Herdeiro:2026sur}.

In this study, we extend this phase transition view to two different directions. We investigate the effect of a linear term in the scalar field coupling to matter (in the so-called \emph{conformal factor}) and also the effect of rotation in the astrophysical systems. The tachyonic instability is mainly a result of the quadratic term in the scalar field coupling of the scalar-tensor theory, so any lower order terms are ignored in many studies, including the original one by DEF~\cite{Damour:1993hw}. However, one would expect to have a linear term in the generic case, hence, it is natural to investigate this for a thorough understanding of scalarization. Rotation is also a generic property of compact objects, and it is known to have drastic effects on the characteristics of scalarization \cite{Doneva:2013qva}, hence it is another avenue for exploring the phase transition. The two avenues we explore here are far from straightforward extensions of what has been achieved. Rather, they each have particular relevance to understanding scalarization as a phase transition. 

For the linear coupling, we will see that the phenomenological understanding of the phase transition view provides simple yet very powerful prediction tools for what sort of solutions are possible in the scalar-tensor theory. We will demonstrate that the basic structure of the Landau energy ansatz allows us to determine the number of solutions and their basic properties like stability or scalar field amplitudes. We will present how the previous phenomenology, which lacked a linear coupling, is modified, and how the predictions of the new picture are fully borne out in our numerical results. For example, we will see that the phase transition picture enables us to find certain branches of scalarized solutions which are commonly missed in numerical searches. 

Previous studies demonstrated that rotation of neutron stars enhances the maximum possible scalarization strength and expands the range of parameters where neutron stars with scalar hair can exist \cite{Doneva:2013qva,Doneva:2016xmf}. This is also relevant for the newly discovered prominence of first-order scalarization. In the known examples so far, even though first-order scalarization is very common in the parameter space, the interesting stellar mass range where metastable states and possible transitions between them are possible, is below one solar mass. This makes these theoretical results of limited observational relevance, since most observed neutron stars have higher masses \cite{Ozel:2016oaf}. Hence, it is interesting to check whether rotation can increase the first-order phase transition mass. We will show that indeed this critical mass increases, however the magnitude of the change is not sufficient to make the solutions substantially more interesting for astrophysics. 

The paper is structured as follows: In Section~\ref{sec: st theory} we set the mathematical foundation of the studied scalar-tensor theory. In Section~\ref{sec:phase transition} we review the phenomenological analysis of the Landau type for the pure DEF theory, and adapt it for the presence of a linear term in the coupling.  In Section~\ref{sec: results} we present the numerical results for the phase transition picture, examine the full spectrum of solutions in the theory, and compare these to the predictions from Section~\ref{sec:phase transition}. The paper ends with our conclusions. 

\section{Scalar-tensor theory}
\label{sec: st theory}

We will work within scalar-tensor theories defined by the following form of the Einstein frame action   
\begin{eqnarray}
S&=&\frac{1}{16\pi G_{*}}\int d^4x \sqrt{-g} \left(R -
2g^{\mu\nu}\partial_{\mu}\varphi \partial_{\nu}\varphi -
4V(\varphi)\right) \nonumber\\ && + S_{m}[\Psi_{m}; {\cal A}^{2}(\varphi)g_{\mu\nu}],
\end{eqnarray}
where $G_*$ is the bare gravitational constant, $R$ is the Ricci scalar with respect to the Einstein frame metric $g_{\mu\nu}$, and $V(\varphi)$ is the scalar field potential. ${\cal A}(\varphi)$ is the conformal factor relating the Einstein frame and the physical Jordan frame metrics, $g_{\mu\nu}$ and ${\tilde g}_{\mu\nu}$ respectively, namely ${\tilde g}_{\mu\nu}={\cal A}^{2}(\varphi) g_{\mu\nu}$. The matter fields are collectively denoted by $\Psi_{m}$.  Note that the action and the field equations below are given in the Einstein frame for mathematical convenience while the physical quantities presented in the figures are transformed back to the physical Jordan frame.

The Einstein frame field equations obtained by varying the above action are
\begin{eqnarray} 
R_{\mu\nu} - \frac{1}{2}g_{\mu\nu}R &=& 8\pi G_{*} T_{\mu\nu}
+ 2\partial_{\mu}\varphi \partial_{\nu}\varphi   -  \label{EFFE1}\\ 
&& g_{\mu\nu}g^{\alpha\beta}\partial_{\alpha}\varphi
\partial_{\beta}\varphi -2V(\varphi)g_{\mu\nu}  \,\,\, , \nonumber \\
\nabla^{\mu}\nabla_{\mu}\varphi &=& - 4\pi G_{*} k(\varphi)T
+ {dV(\varphi)\over d\varphi} , \label{EFFE}
\end{eqnarray}
where $T_{\mu\nu}$ is the Einstein frame energy-momentum tensor, $T$ is its trace and $k(\varphi)$ is the logarithmic derivative of the conformal factor ${\cal A}(\varphi)$  defined as
\begin{equation}
k(\varphi)= \frac{d\ln({\cal  A}(\varphi))} {d\varphi}.
\end{equation}

We will consider a scalar field with mass $m_\varphi$ where the potential is given by
\begin{equation} \label{eq:pot}
    V(\varphi) = \frac{1}{2}m_{\varphi}^2\varphi^2 , \quad
\end{equation}
but also study the $m_\varphi=0$ case. The conformal factor is chosen to be a combination of the Brans-Dicke (BD) theory with a constant $k(\varphi)$ and the  DEF model with linear $k(\varphi)$, namely 
\begin{equation} \label{eq:coup}
    k(\varphi) = \alpha + \beta \varphi,
\end{equation}
where $\beta$ and $\alpha$ are constants. Thus ${\cal A}(\varphi) = e^{\alpha \varphi + \frac{1}{2}\beta \varphi^2}$.

The pure DEF model with $k(\varphi) = \beta \varphi$ allows for the so-called spontaneous scalarization which is discussed in detail in the following section. It is straightforward to notice that such a coupling function keeps the field equations \eqref{EFFE1}, \eqref{EFFE} invariant with respect to a sign change of $\varphi$. The BD linear term $k(\varphi)=\alpha$, though, breaks the symmetry with respect of the scalar field sign. Interestingly, due to this term, an extra polarization of the gravitational waves appears, the so-called breathing modes. Thus, radial perturbations of a compact object lead to gravitational wave emission.

In the present study, we are interested in both static and rapidly rotating neutron star solutions. For the neutron star matter, we assume a perfect fluid with an energy-momentum tensor being
\begin{eqnarray}
T_{\mu\nu}= (\rho + p)u_{\mu} u_{\nu} + pg_{\mu\nu},
\end{eqnarray}
where $p$, $\rho$ and $u_{\mu}$ are the Einstein frame pressure, energy density and four velocity of the fluid, respectively. The Einstein frame energy-momentum tensor conservation takes the following form
\begin{eqnarray}
\nabla_{\mu}T^{\mu}{}_{\nu} = k(\varphi)T\partial_{\nu}\varphi .
\end{eqnarray}
The fluid four velocity for an axisymmetric star can be written as
\begin{equation}
u^\mu = \frac{e^{-(\sigma + \gamma)/2}}{\sqrt{1-v^2}}
[1,0,0,\Omega], \label{eq:four_velocity}
\end{equation}
where $\Omega=\frac{u^{\phi}}{u^{t}}$ is the angular velocity and the proper velocity $v$ of the fluid is given by $v = (\Omega - \omega) r \sin \theta e^{-\sigma}$. In the case of uniform rotation we study here, the angular velocity is a constant throughout the star. 

Let us comment on the relation between the Einstein and the physical Jordan frames for some of the most important quantities. The energy-momentum tensor, the energy density, the pressure and the four velocity are transformed between the two frames as
\begin{eqnarray}\label{DPTEJF}
T_{\mu\nu} &=& {\cal A}^2(\varphi){\tilde T}_{\mu\nu}, \nonumber \\
\varepsilon &=&{\cal A}^4(\varphi){\tilde\varepsilon}, \nonumber \\
p&=&{\cal A}^4(\varphi){\tilde p},  \\
u_{\mu}&=& {\cal A}^{-1}(\varphi){\tilde u}_{\mu}, \nonumber
\end{eqnarray}
where the Jordan frame quantities are denoted with a tilde. The angular velocity $\Omega$, the proper velocity $v$ and the angular momentum $J$ remain the same in both frames. The definition of gravitational (Arnowitt-Deser-Misner) mass is somewhat subtle in scalar tensor theories (see e.g. \cite{Doneva:2013qva}) but for the considered coupling function it coincides between the two frames.

In accordance with the physical model at hand we consider stationary and axisymmetric matter and scalar field configurations with the following Einstein frame ansatz for the line element in quasi-isotropic coordinates
\begin{eqnarray} \label{eq:metric}
ds^2 &=& -e^{\gamma+\sigma} dt^2 + e^{\gamma-\sigma} r^2
\sin^2\theta (d\phi - \omega dt)^2 + \\  &&e^{2\eta}(dr^2 + r^2
d\theta^2) \nonumber,
\end{eqnarray}
where  the metric functions depend only on  $r$ and $\theta$. The dimensionally reduced field equations can be derived after substituting the above metric in the field equations \eqref{EFFE1}, \eqref{EFFE}. Since they are quite lengthy, we refer the interested reader to \cite{Doneva:2013qva, Doneva:2016xmf, Yazadjiev:2015zia} for their explicit form.

For the line element \eqref{eq:metric}, the angular momentum $J$, the Arnowitt-Deser-Misner (ADM) and the baryon masses are defined by the following expression in terms of the Einstein frame quantities:
\begin{equation}
J = \int_{\mathrm{Star}} (\varepsilon + p)\,\frac{v^2}{1 - v^2}\,\frac{1}{\Omega - \omega}\,\sqrt{-g}\, d^3 x,
\end{equation}

\begin{eqnarray}
    M_{\textrm{ADM}} &=& \int_{\mathrm{Star}} \left[
(\varepsilon + 3p) + \right. \\ &&\left. 2(\varepsilon + p)\,\frac{\Omega}{\Omega - \omega}\,\frac{v^2}{1 - v^2}
- \frac{1}{2\pi} V(\varphi)
\right] \sqrt{-g}\, d^3 x \nonumber
\end{eqnarray}
and 
\begin{equation}
    M_{\textrm{b}} = \int_{\mathrm{Star}} A^3(\varphi)\,\tilde{\rho}\,u^t \sqrt{-g}\, d^3 x,
\end{equation}
where $\tilde{\rho}$ is the rest mass density in Jordan frame. In the next sections, we will present the angular momentum in geometrical units $J = \frac{Jc}{GM^{2}_{\odot}}$.

 The physical circumferential radius of the star in Jordan frame differs from the Einstein frame one and it is
\begin{equation}
 R_e = {\cal A}(\varphi)\; r \;
e^{(\gamma-\sigma)/2}|_{r=r_{e},\theta=\pi/2},
\end{equation}
where $r_e$ is the Einstein frame coordinate equatorial radius of the star, defined as the point where the pressure vanishes.

\section{Spontaneous scalarization and gravitational phase transitions}
\label{sec:phase transition}

\subsection{Current understanding with $\alpha =0$}    

Spontaneous scalarization occurs when the scalar field carries an unstable mode that is triggered in certain astrophysical environments, such as neutron stars above a certain mass~\cite{Doneva:2022ewd}. This so-called tachyonic instability is eventually suppressed by nonlinear terms, and the final stable configuration is an object that is dressed with a scalar field. This phenomenon can also be understood as a phase transition, which is our main point of view in this paper. Even though scalarization has been known to be a form of phase transition since early days~\cite{Damour:1996ke}, this approach only recently came to the forefront~\cite{Unluturk:2025zie,Muniz:2025egq}. 

Let us first summarize the current understanding of the scalarization phase transition for the special case of $\alpha=0$ in Eq.~\eqref{eq:coup}. We will be using the phenomenological approach of Landau~\cite{Plischke2006, Goldenfeld2018}, mainly following ~\textcite{Unluturk:2025zie}. We determine the equilibrium solutions based on the total energy of the astrophysical system, the  ADM mass $M_\text{ADM}$. In the phenomenological approach, we propose an \emph{ansatz}, a function that depends on the attributes of the system~\cite{Plischke2006,Goldenfeld2018}
\begin{eqnarray} \label{eq: energy landau ansatz}
    M_\text{ADM}(M_\text{b},Q) &=& M_0(M_\text{b}) + a(M_\text{b}) Q^2 + \\ &&\frac{1}{2} b(M_\text{b}) Q^4 +  \frac{1}{3} c(M_\text{b}) Q^6 . \nonumber
\end{eqnarray}
Here, $M_\text{b}$ is the baryon mass of the system, which is the total mass of the constituents of the star if we moved them far away from each other. $Q$ is a measure of the ``strength of scalarization,'' whose precise meaning is not essential. One can use the scalar charge for massless scalars, or the value of the scalar field at the center of the star in general as $Q$~\cite{Unluturk:2025zie,Muniz:2025egq}.\footnote{The scalar charge~\cite{Damour:1993hw} is only well defined for massless and non-self-interacting scalars. Hence, here we will mainly consider the value of the scalar field at the center of the star as the order parameter for uniformity between massive and massless scalars.} This formula can be interpreted as trying to understand what happens if we are given some fixed number of neutrons (corresponding to an $M_\text{b}$ value) and try to build a star with them while also dressing the star with some scalar field cloud. Is the star going to be scalarized or not? If it is scalarized, how large is the scalar field amplitude? We posit that the total energy of any configuration will be a function of how many neutrons we have and how much scalar field we add on top. Then, for a given $M_\text{b}$, which acts similarly to temperature in thermodynamic systems, the equilibrium configurations are the one(s) that are the extremum points of the total energy $M_\text{ADM}(M_\text{b},Q)$ as a function of $Q$. 

How did we pick the specific form of $M_\text{ADM}(M_\text{b},Q)$? It might look like we have some strong assumptions since our ansatz~\eqref{eq: energy landau ansatz} is in a specific form, however this is not the case. If we assume that $M_\text{ADM}$ is an analytic function of its arguments, then the above functional form is generic as long as $Q$ is close to zero so that higher terms in the series expansion can be ignored. We want $c>0$ to ensure that the energy is bounded from below, and there are only even powers for the case at hand due to the $\varphi \to -\varphi$ symmetry of the theory when $\alpha=0$.

\begin{figure}
    \centering
    \includegraphics[width=\columnwidth]{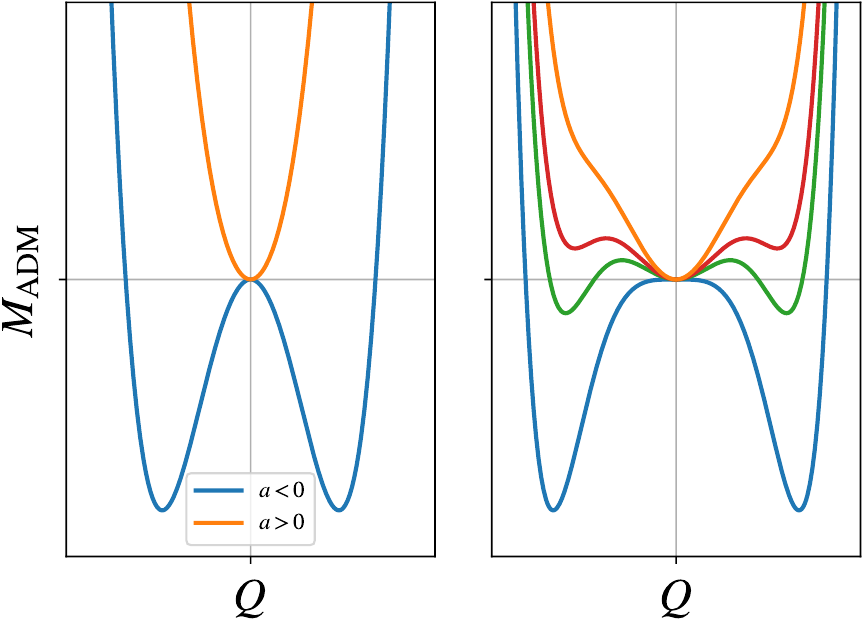}
    \caption{Scalarization as a phase transition. Different curves represent the Landau Ansatz~\eqref{eq: energy landau ansatz} for different values of the baryon mass, lower curves corresponding to larger $\Mb$. Left: A second-order phase transition which occurs when $a(M_\text{b})$ changes sign ($b(M_\text{b}),c(M_\text{b})>0$ in all cases). The sign of $a(M_\text{b})$ determines whether the unscalarized $Q=0$ solution is the only equilibrium, or we also have two symmetric scalarized $Q \neq 0$ configurations. Baryon mass $M_\text{b}$ of the star acts similar to temperature, and $Q$ has the role of an order parameter, like magnetization in spontaneous magnetization. Right: A first-order phase transition which occurs when $b(M_\text{b})<0$ ($c(M_\text{b})>0$ in all cases). As the value of $a(M_\text{b})$ moves from negative to more positive, we move to the higher curves. The intermediate curves predict metastability: both unscalarized and scalarized solutions can be locally stable, and the astrophysical environment can potentially pick either one. Scalarization has been almost exclusively considered to be a second-order transition until recently, but it is now known that the first-order transition is quite common in the parameter space~\cite{Unluturk:2025zie}. }
    \label{fig: landau energy}
\end{figure}
The two main forms of the scalarization phase transition can be schematically seen in Fig.~\ref{fig: landau energy}. Let's start with the case of $b>0$ (left panel) in Eq.~\eqref{eq: energy landau ansatz}. The $c$ term does not play an essential role in this case, since it is sufficient to concentrate on a small region around $Q=0$ where the $Q^6$ term is ignorable. We have a single stable equilibrium at $Q=0$ if $a>0$, which is the case of no scalarization, where we only have the GR solution. If $a<0$, we have 3 equilibrium solutions (extrema), and $Q=0$ is still one of these. However, $Q=0$ is now an unstable solution due to the tachyonic instability we explained above. We have two symmetric stable equilibria at $Q = \pm \sqrt{-a/b}$, and the star assumes one of these configurations depending on the initial and boundary conditions. Scalarization is continuous in the sense that we have stable scalarized stars when $a<0$, and their strength of scalarization, $Q$, can be arbitrarily small in the $a \to 0$ limit. This is a typical \emph{second-order phase transition} that occurs at $a(M_\text{b})=0$, and a direct analog of spontaneous magnetization~\cite{Plischke2006, Goldenfeld2018}. $Q$ acts as magnetization and $M_\text{b}$ acts as temperature.

When $b<0$, possibilities increase as in the right panel of Fig.~\ref{fig: landau energy}. When $a<0$, the case is similar to the second-order phase transition with one unstable and two stable equilibrium configurations (the lowermost curve). We also have a unique stable equilibrium at $Q=0$ for \emph{large} positive $a$, which is the case of no scalarization, again similarly to the second-order transition (the uppermost curve). However, for lower positive values of $a$, there is a new regime. We have three \emph{locally} stable equilibria (minima), and two unstable ones (maxima) (the intermediate curves). Depending on the value of $a$, the solution at $Q=0$ or the scalarized ones with $Q \neq 0$ might be the absolute ``ground state'' of the system (the other being metastable), but all are astrophysically viable solutions since they are stable to small enough perturbations. As for the global energy minimum, it \emph{discontinuously} jumps from $Q=0$ to $Q \neq 0$ at some baryon mass value between the two intermediate curves in Fig.~\ref{fig: landau energy}.\footnote{This is exactly the point where the phase transition is commonly said to occur, that is, it is based on the change in the global minimum. Note that the appearance of the locally stable states at $Q \neq 0$ is before this point (the red line in Fig.~\ref{fig: landau energy}), so the appearance of more than one astrophysically relevant configurations happens at a different $\Mb$ than where the phase transition is in this nomenclature.} This is a typical \emph{first-order phase transition}.

Until recently, the spontaneous scalarization has mostly been considered as a second-order phase transition with very few exceptions~\cite{Kuan:2022oxs}. However, recent work has shown that, the picture can be different at least for the original scalarization model of DEF~\cite{Damour:1993hw} and its massive extensions~\cite{Ramazanoglu:2016kul,Yazadjiev:2016pcb}: first-order transitions dominate the $(\beta, m_\varphi)$ parameter space of the theory, and second-order scalarization is relatively exceptional when all theory parameter values are considered~\cite{Unluturk:2025zie}. This is an important point, since first-order scalarization provides a different solution space. For example, it allows metastable solutions that might be relatively easily excited, which can lead to novel observational signatures. 

\subsection{Phenomenological expectations for $\alpha \neq 0$}    
\label{sec: alpha}

In this study, one of our main interests is how the above picture changes when $\alpha \neq 0$. At first, seemingly the only substantial change is that we now have the odd $Q$ powers in the Landau ansatz, Eq.~\eqref{eq: energy landau ansatz}, since the $\varphi \to -\varphi$ symmetry is broken. This might look trivial, yet, it has interesting consequences. We will consider adding a single linear term in $Q$ to the Landau Ansatz for the sake of demonstration, and our exact numerical results will establish that this is sufficient to explain the main qualitative aspects of the $\alpha \neq 0$ case.

\begin{figure}
    \centering
    \includegraphics[width=\columnwidth]{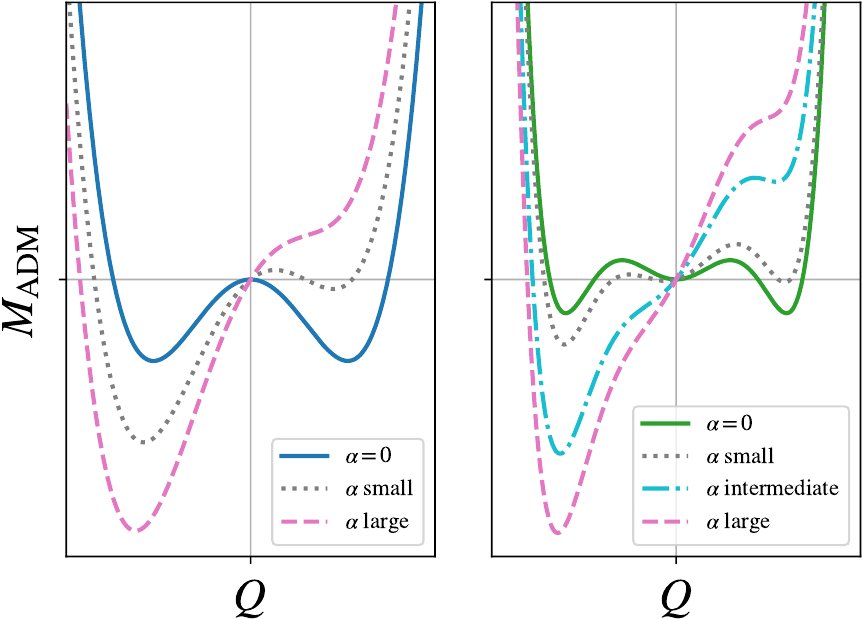}
    \caption{The effect of nonzero $\alpha$ on the phase transition picture. Left: The blue $\alpha=0$ curve is the same as the one in the left panel of Fig.~\ref{fig: landau energy}. As we increase $\alpha>0$, the two local minima become asymmetric, and the (unstable) unscalarized solution at $Q=0$ is modified into a slightly scalarized one. One of the minima is eventually lost for large enough $\alpha$, leading to a radical change in the structure of the scalarized solution branches of the theory. Right: The green curve is the same as the one in the right panel of Fig.~\ref{fig: landau energy}. The number of equilibrium solutions changes from 5 to 3, and then to 1 as we increase $\alpha$. This decrease in the number of solutions is due to the fact that we lose the unstable equilibria with increasing $\alpha$, ending up with a case where there is no phase transition at all for sufficiently large $\alpha$ values, that is, there is only a single stable equilibrium configuration for any value of $\Mb$. This only remaining phase is a strongly scalarized one, resembling the solutions to the Brans-Dicke theory.  }
    \label{fig: landau energy alpha}
\end{figure}

Let us discuss the second-order phase transition first in the left panel of Fig.~\ref{fig: landau energy alpha}, where we look at how increasing $\alpha$ changes the double-welled blue curve in Fig.~\ref{fig: landau energy} which contained the scalarized solutions. For low values of $|\alpha|$, we have small corrections to the $\alpha=0$ picture (the middle curve). $Q=0$ is not an equilibrium configuration anymore, but there is a slightly negatively scalarized local maximum, which is a deformed version of the $Q=0$ GR solution for $\alpha=0$. More generally, there is no unscalarized solution when $\alpha \neq 0$, unlike the previous subsection, as can be surmised from the field equations.
Out of the two local minima, the analogs of the symmetric scalarized solutions at $Q=\pm \sqrt{-a/b}$ for $\alpha=0$, one becomes a deeper minima while the other becomes shallower. Overall, all three equilibrium solutions we have for $\alpha \neq 0$ so far are slight modifications of the three equilibrium solutions we had for $\alpha=0$.
 
The picture radically changes if we increase $|\alpha|$ further. The trend of making the local minimum on one side shallower eventually leads to its total disappearance, which leads to a change in the number of possible equilibrium configurations. Fig.~\ref{fig: landau energy alpha} clearly shows that we have only one stable equilibrium for large enough $|\alpha|$, which is highly scalarized. For this single solution, the scalar field amplitude has the opposite sign to $\alpha$ for our choice of the coupling function ${\cal A}(\varphi)$. In short, the number of solutions decreases from 3 to 1 in this case.\footnote{We did not show the transformation of the single-welled orange curve in the left panel of Fig.~\ref{fig: landau energy} with increasing $\alpha$ to avoid clutter. It also gets shifted to the left and becomes asymmetric. This curve still contains a unique equilibrium solution for $\alpha \neq 0$ which is still stable, but instead of being an unscalarized one as for $\alpha=0$, it has negative $Q$ now. In short, the number of solutions is 1 irrespective of the value of $\alpha$. }

There is a very similar trend when we start with first-order scalarization at $\alpha=0$, which can be seen in the right panel of Fig.~\ref{fig: landau energy alpha}. This time, we start with a case where there are 5 equilibrium solutions as an example (the $\alpha=0$ curve is the three-welled green one in the right panel of Fig.~\ref{fig: landau energy}).
As we increase $|\alpha|$, this number first goes down to 3, and eventually 1. Hence, when we numerically compute our scalarized neutron star solutions, the behavior of the scalarized and unscalarized branches can change drastically depending on $\alpha$.\footnote{If we started with a curve of 3 equilibrium solutions for $\alpha=0$ (e.g. the blue curve on the right subplot of Fig.~\ref{fig: landau energy}), we would go down to 1 equilibrium with increasing $\alpha$. A case with a single solution for $\alpha=0$ continues to have a single solution when $\alpha \neq 0$.}

Our exposition above might suggest that, for example, having 5 equilibria for a first-order scalarization is the ``standard'' picture, and having three of them or a single one for large $|\alpha|$ is the ``surprise.'' However, the case can be just the opposite. Namely, it is not a priori clear how many scalarized or unscalarized (or very slightly scalarized) solutions one would have for a given $M_\text{b}$ if our phase transition approach is not employed. Hence, many studies in the literature were unaware of the richness of the equilibrium solution structure in spontaneous scalarization models. For example, when one mainly works with a large $|\alpha|$ that contains a single highly scalarized branch of solutions as in Fig.~\ref{fig: landau energy alpha}, it is quite easy to completely miss the fact that there are other physical configurations when $|\alpha|$ is smaller. In this sense, a phase transition-based understanding facilitates a systematic exploration of spontaneous scalarization. We will confirm these phenomenological discoveries with concrete numerical solutions in the next section.

Aside from $\alpha \neq 0$, we are also interested in the effect of rotation, $\Omega \neq 0$, on the phase transition behavior. However, the relationship of rotation to our energy ansatz is more opaque than that of $\alpha$. Hence, we could not provide clear predictions for our later numerical computations. Hence, a deeper understanding of the scalarization phase transition in rotating stars awaits future studies.

We will concentrate on the phase transition due to the first appearance of scalarization at lower stellar masses, similarly to \textcite{Unluturk:2025zie}. However, note that for some STT parameters, there can be a second, similar phase transition at higher stellar masses where the scalarized branch rejoins the branch of the GR solutions~\cite{Kuan:2022oxs,Muniz:2025egq}. We will not have a separate discussion, but this phenomenon can also be studied with the methods we present here.

Finally, we should add that the phenomenological understanding provides a more qualitative picture than a quantitative one, so exact numerical details of the solutions cannot always be obtained by a low-order polynomial~\cite{Muniz:2025egq}. This is especially the case for first order scalarization where the solutions at the local minima away from $Q=0$ cannot be fully captured in a small $Q$ expansion, since they appear discontinuosly and away from the origin.

In the following discussion, we will refer to the cases in the left panel of Fig.~\ref{fig: landau energy alpha} as \emph{deformed second-order scalarization}, and the right panel as \emph{deformed first-order scalarization.} We will also use the number of equilibrium neutron star solutions to differentiate different cases.

\section{Results}
\label{sec: results}

\subsection {Computational setup}

The last remaining ingredient to be determined after Section~\ref{sec: st theory} in order to obtain stellar solutions is the nuclear matter equation of state (EOS). The qualitative features of the first-order phase transition observed in STT by~\textcite{Unluturk:2025zie} were not sensitive to the employed EOS. Since the main purpose of the present paper is to qualitatively study the effect of rotation and the linear term $\alpha$, we employ a single realistic EOS, MPA1~\cite{Muther:1987xaa}, in both tabulated and  piecewise polytropic  approximation forms~\cite{Read:2008iy}.

For constructing the rotating equilibrium solutions presented in the rest of the paper we use a modification of the original $\tt RNS$ \cite{1995ApJ...444..306S} code, which is based on the Komatsu-Eriguchi-Hachisu (KEH) scheme~\cite{Komatsu:1989zz, 1989MNRAS.239..153K} with the transformations introduced by Cook et al.~\cite{1994ApJ...422..227C}. The $\tt {RNS}$ code was first extended to scalar-tensor theories in \textcite{Doneva:2013qva} and later versions included a nonzero scalar field mass \cite{Staykov:2023ose,Doneva:2016xmf} which we employ for the numerical calculations. Due to the numerical difficulties and significantly increased computational cost for obtaining solutions with high scalar field mass and theory coupling parameters, it is useful to employ a 1D code for the static neutron star models based on a shooting method \cite{Yazadjiev:2016pcb,Staykov:2018hhc}. The interplay between the two codes (at least in the static case) allows us to conduct a more precise search for the different types of solutions discussed in Sec.~\ref{sec: alpha}. 

As an additional verification of the $\tt RNS$ code, which is known to have accuracy limitations for high values of $|\beta|$ and high masses of the scalar field \cite{Doneva:2016xmf}, we compared static solutions generated with {\tt RNS} to solutions generated with a modified version of the relaxation-based~\cite{Tuna:2022qqr} code in \textcite{Unluturk:2025zie}. With the increase of $|\beta|$ and $m_{\varphi}$ the global parameters of the solutions (like the gravitational mass, the baryon mass, the radius etc.) reach up to 2\%  difference between the two codes for the most extreme values of the theory parameters we considered. Some quantities, such as the binding energy of the star, are more sensitive to numerical errors, but these are not central to our analysis, hence the {\tt RNS} code is adequate for our purposes.  

\subsection {Non-rotating neutron stars with $\alpha \neq 0$}

\begin{figure*}[]
    \includegraphics[width=0.95\textwidth]{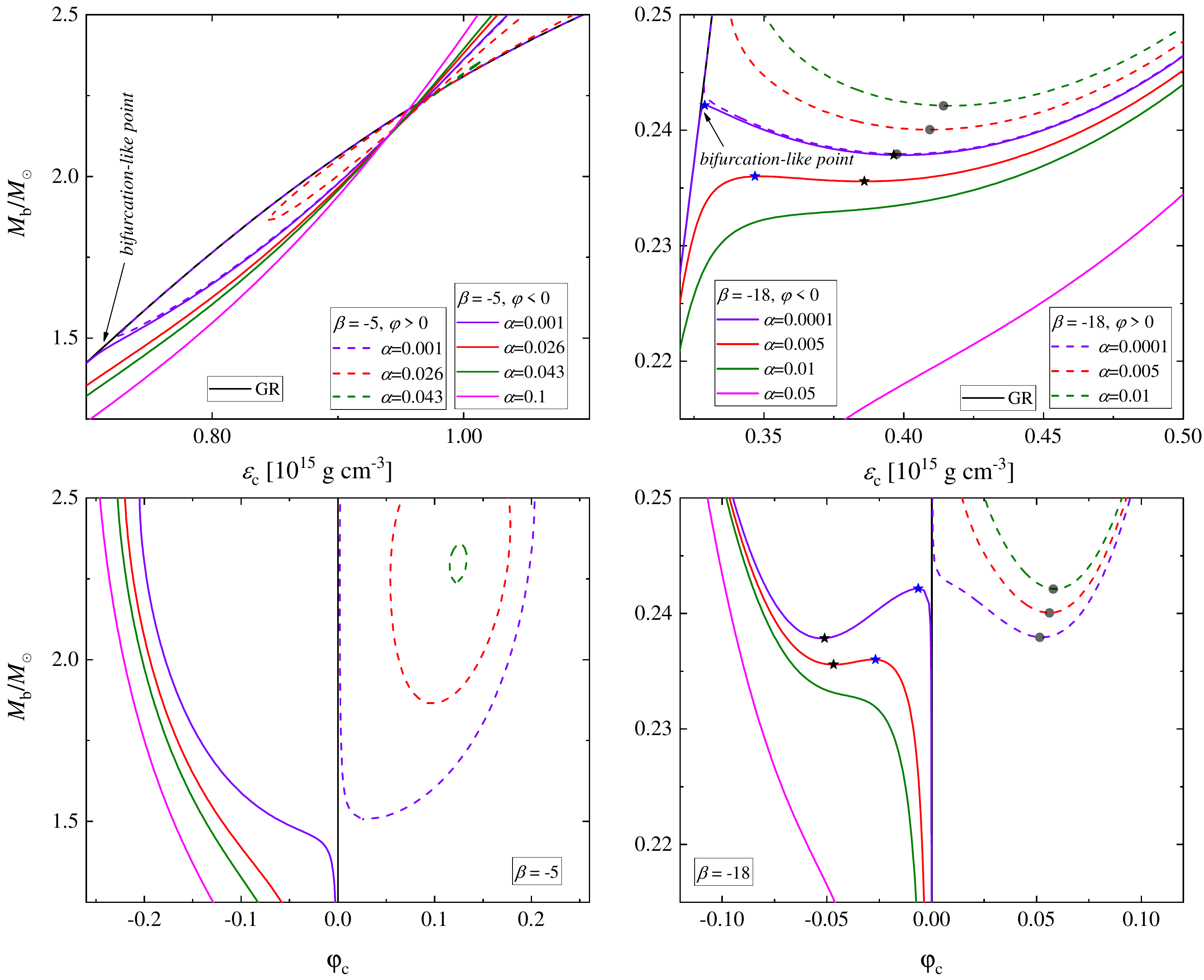}
	\caption{ Static solutions with massless scalar field, $m_\varphi=0$, and different combinations for the parameters $(\beta,\alpha)$. The asterisks (circles) mark the change in stability (coinciding with the mass extrema) on a solution branch with negative (positive) $\varphi$. \textit{Top row:}  The baryon mass $M_{\textrm{b}}$ as a function of the central energy density and \textit{bottom row:} central value of the scalar field as a function of the central energy density. The values for $\beta$ are chosen so that the \textit{left} column represents a deformed second-order phase transition (left panel of Fig. \ref{fig: landau energy alpha})  and the \textit{right} column a deformed first-order phase transition (right panel of Fig. \ref{fig: landau energy alpha}).}
	\label{fig:Mb vs rhoc alpha}
\end{figure*}

In this section, we will focus on spherically symmetric static neutron stars. The scalarized star solutions form a one parameter family with respect to their central energy density, $\varepsilon_c=\varepsilon(r=0)$, hence plots of baryon mass $\Mb$ vs $\epsc$ and the central value of the scalar field $\varphi_\textrm{c}$ vs $\epsc$ are ideal to understand different phases of scalarized solutions (recall that $\varphi_\textrm{c}$ can act as the order parameter $Q$). We also provide the solution curves for GR, which are important to understand the effect of $\alpha$ on scalarization. 

Fig.~\ref{fig:Mb vs rhoc alpha} demonstrates the essential aspects of our results, and shows what happens when we deform first and second-order scalarization with increasing $\alpha$ as we discussed in Sec.~\ref{sec: alpha} (the two right and two left panels, respectively). Let us first look at the lines corresponding to the smallest considered $\alpha$, namely $\alpha=0.001$ in the left column and $\alpha=0.0001$ in the right column, and focus only on the negative $\varphi$ scalarized branches (solid curves). The $\Mb(\epsc)$ curve has two distinct behaviors at the bifurcation-like point, i.e., the point where the scalarized branch starts to significantly deviate from the GR solutions.\footnote{When $\alpha=0$, every GR solution is also a solution of the STT with zero scalar field, and there is an exact point where the scalarized solutions appear for the first time and branch off from the GR solutions at a minimum baryon mass. This is the bifurcation point. When $\alpha \neq 0$, there are no unscalarized solutions, hence there is not a single point we can refer to. Nevertheless, for small $\alpha$, there is still a place where the deviation from GR suddenly increases with a small change in $\epsc$ as marked in Fig.~\ref{fig:Mb vs rhoc alpha}. This is the \emph{bifurcation-like point.}}
$\Mb(\varepsilon_c)$ can have a positive slope, as in the top left panel, where we have deformed second-order scalarization, or it can have a negative slope, as in the top right panel, where we have deformed first-order scalarization. In the latter case, the curve descends until reaching some local minimum, past which it starts ascending again. Note that the difference in the slope directly affects how many solutions we have for a given $\Mb$ value, as we will further discuss below. All these points for deformed first- and second-order scalarization closely resemble their $\alpha=0$ cousins~\cite{Unluturk:2025zie}.
 
Still only considering the negatively ($\varphi<0$) scalarized branches, now let us study increasing values of $\alpha$ on the top right panel. The bifurcation becomes less prominent, and the intermediate part of the branch with descending $\Mb(\epsc)$ becomes shorter. For some critical value of $\alpha$, this subbranch fully disappears and the curve becomes monotonically increasing up to the global maximum (e.g. the $\alpha=0.01$ lines in the figure). This is more akin to a pure BD theory rather than the DEF model, which is expected since the linear term $\alpha$ is associated with this theory \cite{Yazadjiev:2016pcb}.

What about the positively ($\varphi>0$) scalarized branches, the dashed curves, we ignored so far? Let us again concentrate on the top right panel of Fig.~\ref{fig:Mb vs rhoc alpha}, the top left panel can also be understood in similar terms. Note that for small $\alpha$, the dashed curves also start deviating from the GR curve at their own bifurcation-like point, and very closely follow the other (solid) branch afterwards. Indeed, in the $\alpha=0$ limit, the solid and dashed curves coincide away from the GR solutions, which are the two solutions related by the $\varphi \to -\varphi$ symmetry (see the $\alpha=0$ cases in Fig.~\ref{fig: landau energy alpha}). Nonzero $\alpha$ breaks this symmetry, but the solutions are still close to each other in terms of baryon and ADM masses when the asymmetry is small. There is also a part of the positively charged branch that closely follows the GR curve, and can be interpreted as a deformed version of it. Note that the GR curve is not deformed into a single scalarized branch in the STT. At baryon masses above the bifurcation-like point, the GR solutions are closely followed by the positively scalarized (dashed) branch, whereas the negatively scalarized (solid) branch follows the GR one below the bifurcation-like point. The above considerations apply to relatively small $\alpha$. Increasing it makes the deformations and asymmetries larger, pushing the dashed branch to ever higher baryon mass values.

How does this behavior of solutions compare to what we predicted in our phenomenological model? We have the more commonly studied deformed second-order scalarization on the left two panels, which typically occurs at lower $|\beta|$ values. These solutions confirm all the expectations from the left subplot of Fig.~\ref{fig: landau energy alpha}. On top left, when $\alpha$ is small, the maximum number of solutions we can have is 3, which, for example, occurs above $\Mb \approx 1.5 M_\odot$ for $\alpha=0.001$ and $\Mb \approx 1.8 M_\odot$ for $\alpha=0.026$.\footnote{The most strongly scalarized branch re-approaches the GR curve at even higher $\Mb$ values, leading to a second phase transition, as we explained at the end of Sec.~\ref{sec: alpha}~\cite{Kuan:2022oxs}. However, we avoided including this part for clarity of presentation.} The number of equilibrium solutions drops to 1 at lower baryon masses when we have the phase transition. Whereas, when $\alpha$ is above a certain value, we always have exactly 1 solution, as can be seen for the $\alpha=0.1$ lines on the same panels.

The same behavior can also be observed in the bottom left panel in terms of $\varphi_\textrm{c}(\epsc)$, which provides further confirmation of our phenomenological approach. For example, 2 of the equilibrium solutions have negative scalar field amplitudes in all cases when there are 3 solutions in total.\footnote{This is, again, because we are considering positive $\alpha$. Negative alpha would lead to 2 solutions with positive scalar field amplitude and 1 with negative.} Note that the $\varphi_\textrm{c}(\epsc)$ curves also reveal further differences between the negatively and positively scalarized branches for any given $\alpha$. The negative one is continuously connected to the deformed GR solutions we mentioned, whereas the other forms a closed loop.\footnote{We also note that these $\varphi_\textrm{c}(\epsc)$ plots are qualitatively very similar to scalar charge vs the central baryon number density plots of \textcite{Muniz:2025egq} for various scalar field values at infinity, what they call $\varphi_0$ ($\varphi_0 \neq 0$ is possible when the scalar is massless). This is not a coincidence, since both $\alpha$ and $\varphi_0$ introduce linear terms to the Landau ansatz, hence lead to similar outcomes, but their physical meanings are distinct.} 

The right two panels of Fig.~\ref{fig:Mb vs rhoc alpha} have a higher $|\beta|$ value leading to deformed first-order scalarization, and confirm the right subplot of Fig.~\ref{fig: landau energy alpha}. The overall structure is quite similar to the previous paragraph, but now we can have up to 5 solutions which can decrease to 3 as $\alpha$ increases. This change in the number of possible solutions is related to whether the $\Mb$ vs $\epsc$ curve goes up or down at the bifurcation-like point, which was the reason for our relatively detailed discussion before. Even though the right panel of Fig.~\ref{fig: landau energy alpha} suggested that we eventually reach a theory where there is only a single phase if $\alpha$ is high enough, we did not encounter this case. The positively scalarized (dashed) branch appears at higher and higher baryon masses with increasing $\alpha$, rendering it potentially irrelevant from an astrophysical perspective. Nevertheless, we still had such a branch for $\alpha = 10$.

We should emphasize at this point that obtaining Fig.~\ref{fig:Mb vs rhoc alpha} with all the branches is quite nontrivial, which might not be obvious since everything closely follows our phenomenological analysis from Section~\ref{sec: alpha}. When the scalarized branches are numerically computed, one often finds a solution and then moves along the branch that contains this solution. Hence, the existence of separate branches can easily be overlooked. Even though finding all branches is not always central to understanding every specific aspect of scalarization, it provides a more complete picture and can have important astrophysical consequences. It was our earlier phenomenological analysis that enabled us to obtain the exact picture, especially for formed firstde-order scalarization. In some cases, we needed to go back and forth between the KEH scheme in the RNS and the standard shooting method in the 1D code to obtain all of the solution branches. Hence, phenomenology can provide criteria about where to stop the solution search by informing us of the allowed number of solutions. 

\begin{figure}[]
    \includegraphics[width=0.45\textwidth]{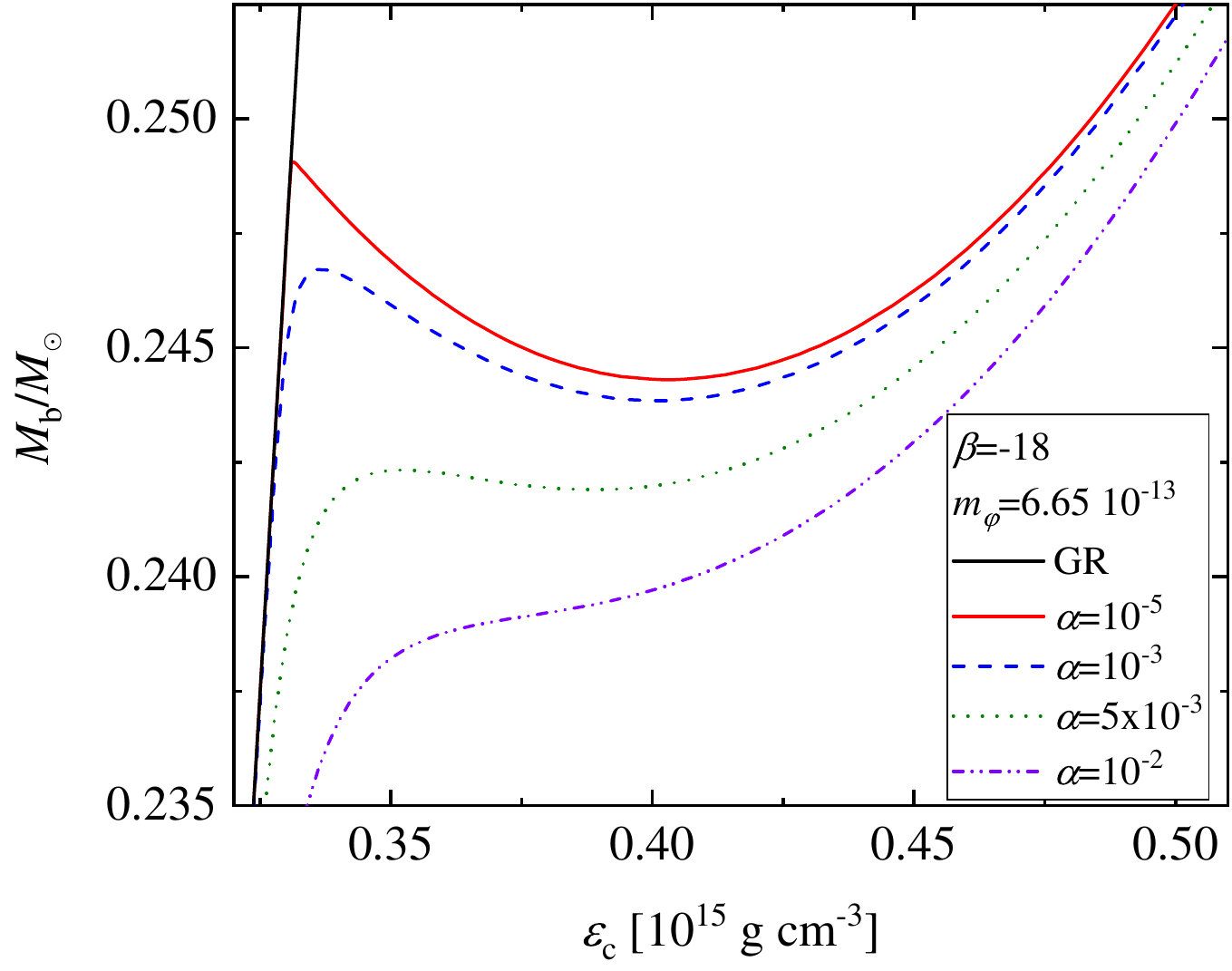}
    \includegraphics[width=0.45\textwidth]{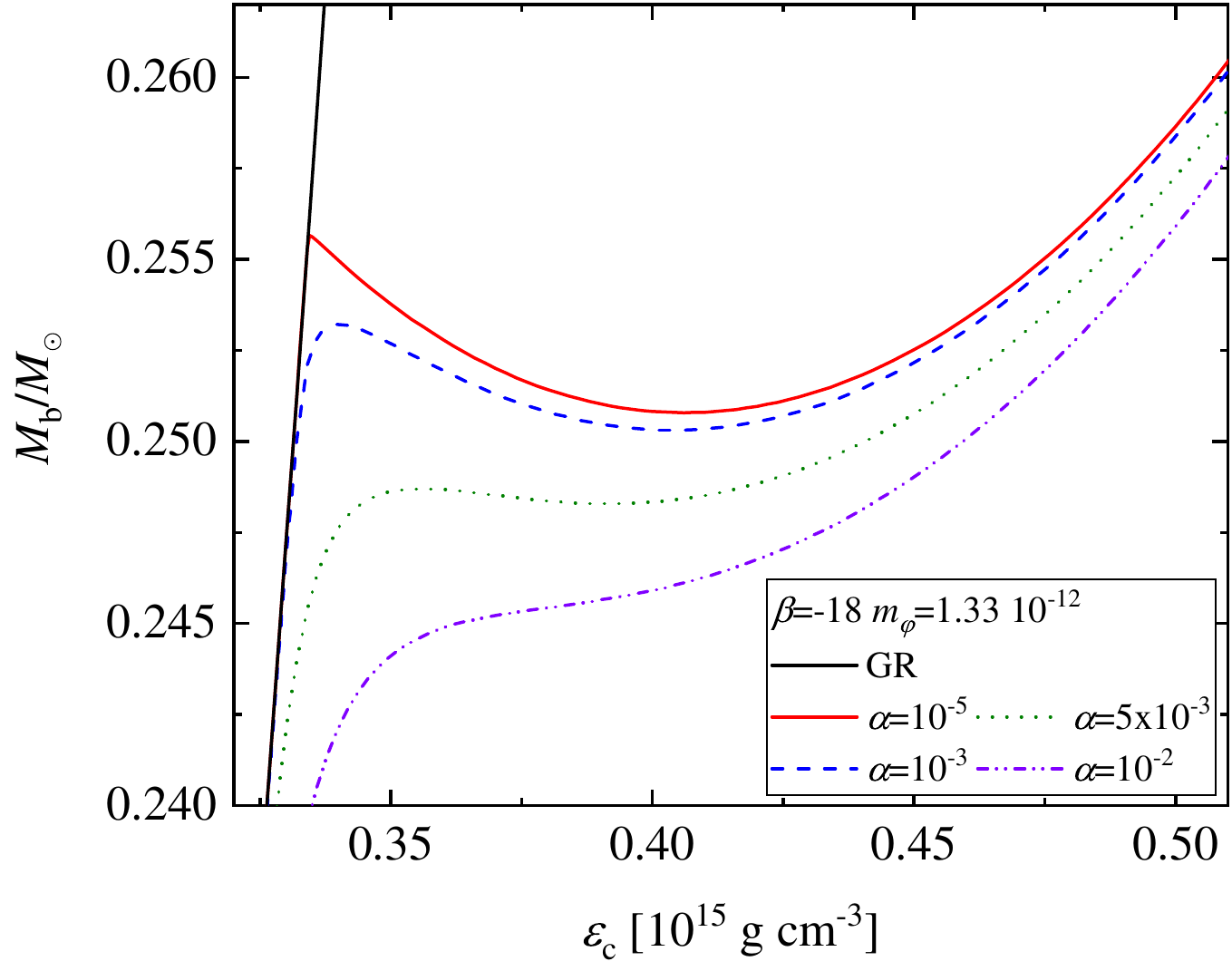}
    \caption{The dependence of $\Mb(\epsc)$ on the scalar field mass: $m_{\varphi}\simeq 6.65 \times 10^{-13}$ eV (\textit{top panel}) and $m_{\varphi}\simeq 1.33 \times 10^{-12}$ eV (\textit{bottom panel}). The coupling parameter is $\beta =-18$ in both cases and we plot several different values of $\alpha$ similarly to Fig.~\ref{fig:Mb vs rhoc alpha}. Increasing $m_{\varphi}$ moves the bifurcation point and the local minimum of $\Mb(\epsc)$ higher, but it also makes the bucket shallower.}
	\label{fig:alpha0_mphi_neq_0}
\end{figure}
Once the basic picture is obtained as above, another question is the dependence on the theory parameters. Which theory parameters lead to deformed first- or second-order scalarization? When we have deformed first-order scalarization, how prominent is the descending branch of $\Mb(\epsc)$, which is a place where additional potentially interesting physics appears? We have already seen the effect of $\alpha$, so the remaining parameter space for our choice of scalar coupling is the $(\beta, m_\varphi)$ plane. \textcite{Unluturk:2025zie} systematically searched the whole $(\beta, m_\varphi, \alpha=0)$ parameter subspace, and showed that  first-order scalarization becomes the norm at smaller $m_\varphi$ and more negative $\beta$.  Together with the fact that scalarization ceases to exist altogether above certain ($\beta$ dependent) $m_\varphi$ values, this indicates that first-order scalarization dominates the parameter plane for $\alpha=0$. 

We observed the effect of the $(\beta, m_\varphi)$ parameters in the case of small nonzero $\alpha$ to be similar to $(\beta, m_\varphi, \alpha=0)$ we just described. To summarize, increasing $m_\varphi$ drives the phase transition from deformed first-order to deformed second-order, and the strongly scalarized solutions completely disappear when $m_\varphi$ gets large enough, only leaving the weakly scalarized solution as the analog of the GR solution for $\alpha=0$. The effect of increasing $\alpha$ for fixed values of $(\beta, m_\varphi)$ also behaves similarly to the case of $m_\varphi=0$: the number of equilibrium solutions decrease to possibly go down to just 1. 

Fig.~\ref{fig:Mb vs rhoc alpha} already indicated that more negative $\beta$ moves us from deformed second-order to deformed first-order scalarization at small $\alpha$ as we described, which is indeed confirmed when a broader range of parameters is explored. Fig.~\ref{fig:alpha0_mphi_neq_0} shows the essential effects of the scalar mass $m_\varphi$ (we only show the $\varphi < 0$ branches, and in different pasterns for increased readability). Increasing the scalar mass makes the ``well'' in the $\Mb(\epsc)$ curves shallower. Ultimately, the well is completely lost at high enough $m_\varphi$, signaling the transition from deformed first- to deformed second-order scalarization.

\subsection{Stability of the scalarized branches}
Lastly, we want to comment on the local and global stability of the solutions we obtained. Recall that not all extrema of the Landau ansatz are stable. The local minima are expected to correspond to (at least) locally stable solutions, whereas local maxima are unstable. We did not perform time evolution computations, which would be the ultimate test for stability, but well-known results about equilibrium solutions of fluids under gravity vindicate the phenomenological picture \cite{Ramazanoglu:2016kul,Mendes:2018qwo}. Consider the $\alpha=10^{-3}$ case in the top right panel of Fig.~\ref{fig:Mb vs rhoc alpha}, which has 5 solutions for $\Mb=0.24 M_\odot$. This baryon mass value corresponds to the dotted gray line on the left subplot of Fig.~\ref{fig: landau energy alpha}, which suggests that two of the solutions are unstable and three of them are locally stable. Moreover, we expect the ground state configuration, the one with the lowest global total energy (ADM mass) also to be the one with the most negative scalar field amplitude (the leftmost minima in Fig.~\ref{fig: landau energy alpha}).

The phenomenological picture is indeed confirmed by the behavior of $\Mb(\epsc)$ in Fig. \ref{fig:Mb vs rhoc alpha}. It is well known that a point where $d\Mb/d\varepsilon_\textrm{c}=0$ separates linearly stable solutions from unstable ones, $d\Mb/d\epsc>0$ being the hydrostatically stable side~\cite{Sorkin:1981jc,Sorkin:1982ut,Shapiro:1983du}. Then, in  the top right panel of Fig.~\ref{fig:Mb vs rhoc alpha}, the descending parts of the curves represent unstable solutions, and the ascending parts stable ones. Among the 5 solutions we mentioned that intersect the $\Mb=0.24 M_\odot$ line, two are unstable and three are stable by this criterion, as expected. This is the baryon mass region with the richest solution structure, since we have only 3 solutions at higher baryon masses, such as $\Mb= 0.25 M_\odot$, and only a single, GR-like solution at lower baryon masses, such as $\Mb= 0.23 M_\odot$ and below.

\begin{figure}[]
    \includegraphics[width=0.45\textwidth]{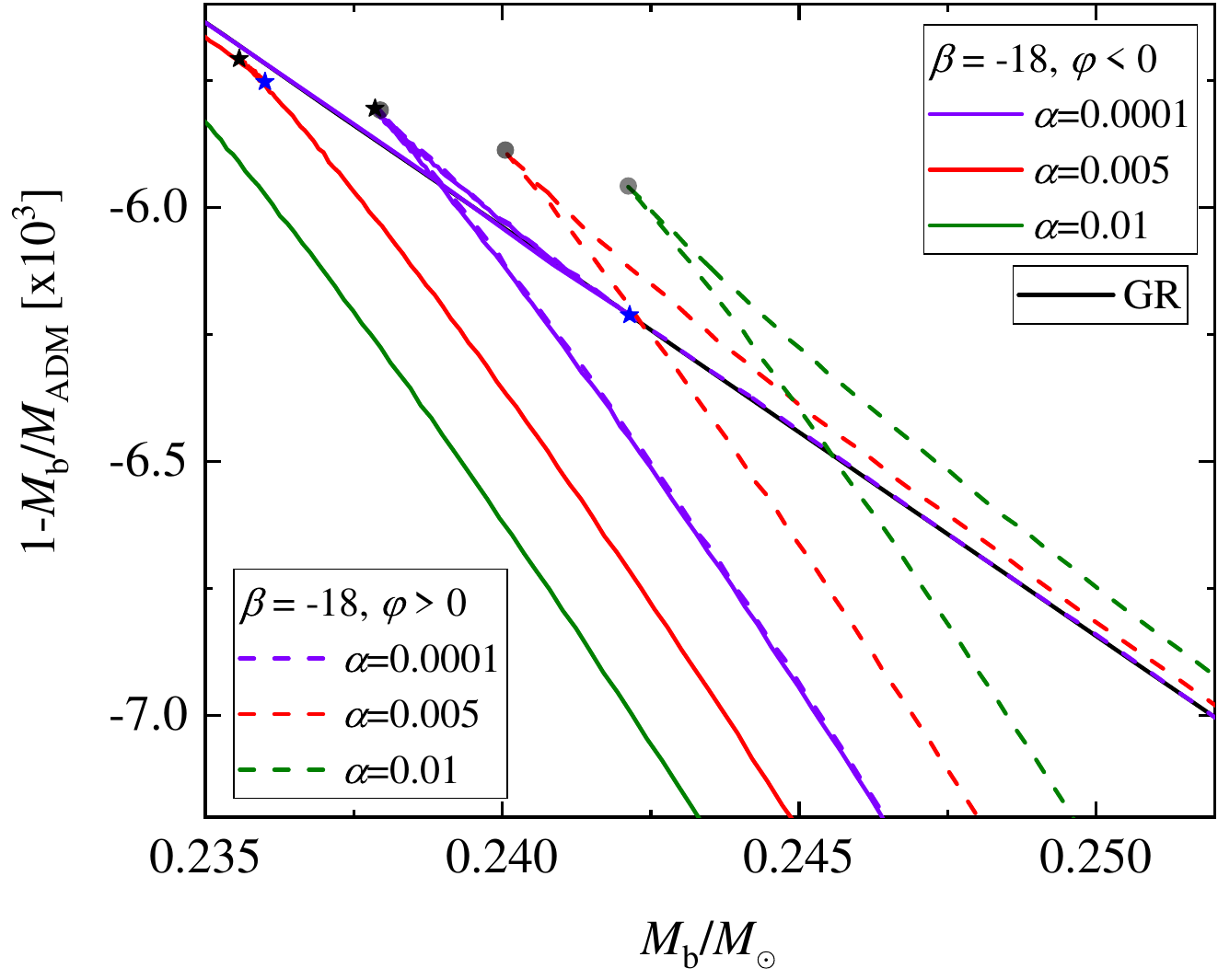}
    \caption{Binding energies of the equilibrium solutions from the right panels of Fig.~\ref{fig:Mb vs rhoc alpha} ($\beta =-18$, $m_{\varphi}=0$). 
    $\alpha= 0.0001$: We reach a cusp where the curve changes direction towards the left, which is exactly the local maximum of the $\Mb$ vs $\epsc$ curve near the bifurcation-like point (Fig.~\ref{fig:Mb vs rhoc alpha} top right). We then encounter a second cusp as we move leftwards, which is the local minimum of the $\Mb$ vs $\epsc$ curve. The curve then continues rightward as $\epsc$ increases further. The solutions between the cusps are most likely unstable due to the arguments we elaborate in the text. 
    $\alpha= 0.005$: Only the positively scalarized (dashed) solutions have cusps. 
    $\alpha= 0.01$: Positively scalarized solutions completely disappear, and there are no cusps for the single negatively scalarized branch. The positions of the cusps are marked with stars and circles which also appear also in the left panel of Fig. \ref{fig:Mb vs rhoc alpha}. }
	\label{fig: cusps nonrotating}
\end{figure}
In order to determine the globally stable ground state configuration and the energy differences between metastable states, we need to compare the total energy, the ADM mass, of the solutions with the same baryon mass. This can be deduced from Fig.~\ref{fig: cusps nonrotating} where we report the binding energy, $\Mb-M_\textrm{ADM}$ per baryon mass. Among other things, this figure also demonstrates that the ground state solution, the configuration with the lowest ADM mass, is a negatively scalarized one for any value of $\Mb$. Further comparison to the $\varphi_\textrm{c}$ plots in the bottom right panel of Fig.~\ref{fig:Mb vs rhoc alpha} also shows that this solution is always the most scalarized one in absolute value and has negative scalar fields, which is another clear prediction of the phenomenological analysis in Fig.~\ref{fig: landau energy alpha}. 

Finally, note that the cusps in Fig.~\ref{fig: cusps nonrotating} are hallmarks of a binding energy plot since hydrostatical considerations ensure that the ADM mass and the baryon mass reach their extrema at the same point, that is $d\Mb/d\epsc=0$ whenever $dM_{\mathrm{ADM}}/d\epsc=0$~\cite{Shapiro:1983du}.  For better readability we tagged the cusps in Fig.~\ref{fig: cusps nonrotating} and the corresponding solutions in the top right panel of Fig.~\ref{fig:Mb vs rhoc alpha} with symbols.  The existence of cusps also acts as a numerical test of our results, since errors in the equations or bugs in the code often prevent the ADM and baryon masses from having extrema at the same point.

\subsection {Rotating neutron stars with $\alpha \neq 0$}
%
\begin{figure}[]
    \includegraphics[width=0.45\textwidth]{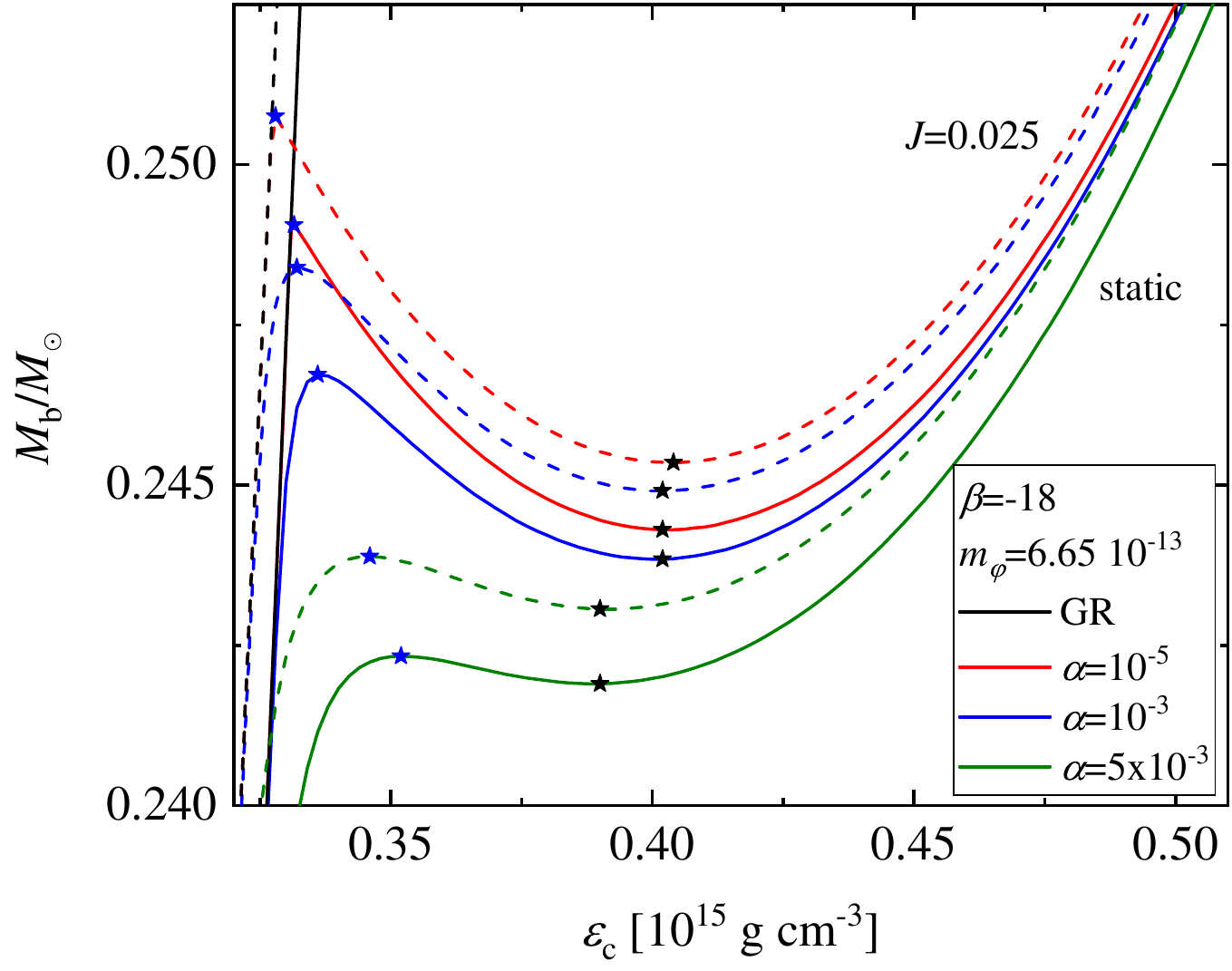}
    \includegraphics[width=0.45\textwidth]{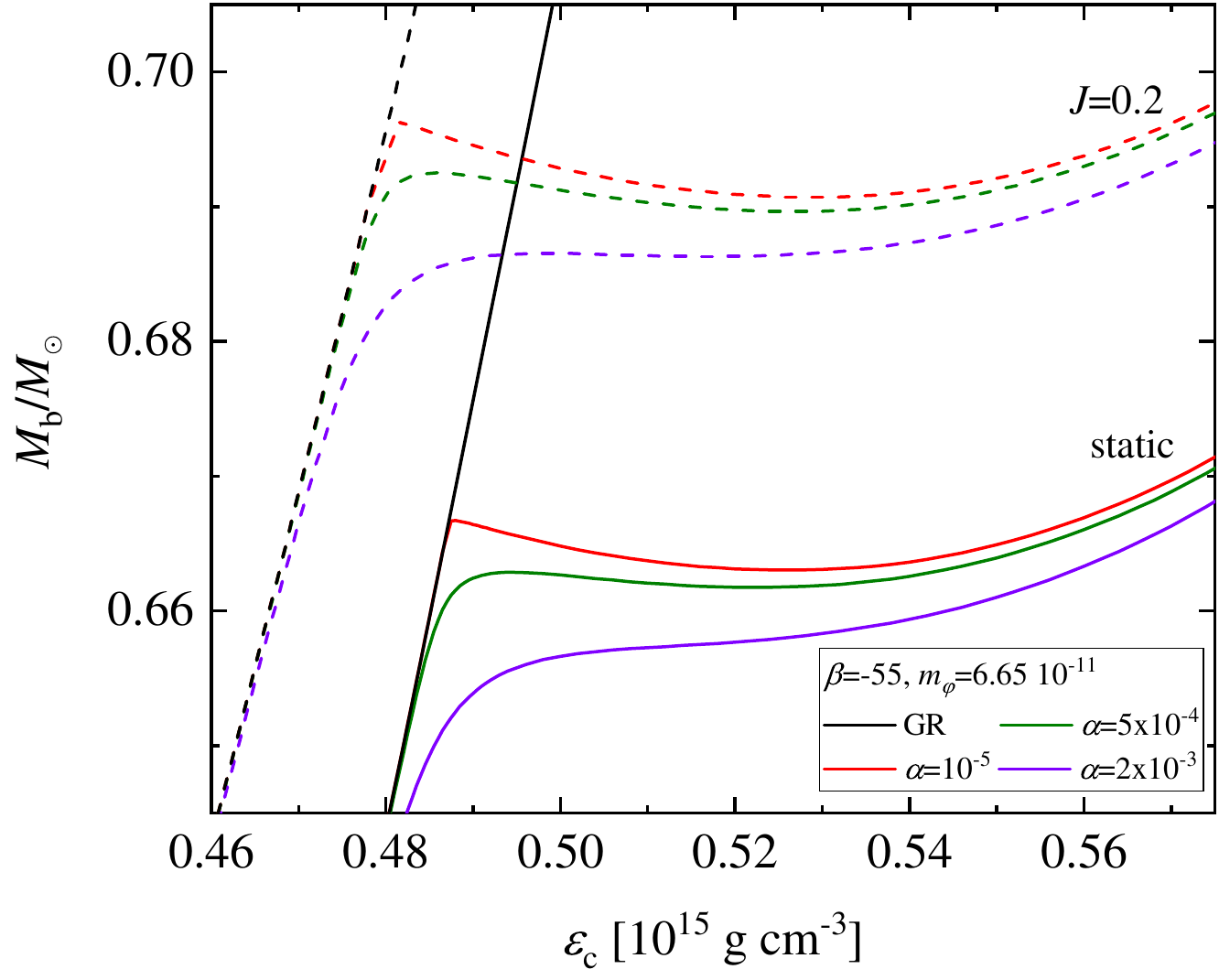}
    \caption{ Comparison between rotating and static solutions. Rotation increases the stellar masses in all cases, and the increase is larger for larger stellar angular momentum. \textit{Top:} $\beta = -18$ and $m_{\varphi}\simeq 6.65 \times 10^{-13}$ eV, where the stellar masses around the bifurcation like point are low and, therefore, only a  lower value of $J=0.025$ is possible before the mass-shedding limit is reached. The small $J$ leads to small changes due to rotation. The stars mark the extrema of the baryon mass. \textit{Bottom:} $\beta = -55$ $m_{\varphi}\simeq 6.65 \times 10^{-11}$ eV where a higher $J$ value is possible due to the increased bifurcation-like mass and there is a correspondingly higher deviation from the non-rotating case. The stellar masses in these plots are lower than those that have been observed so far \cite{Freire:2024adf,Manchester2005ATNF,ATNFOnline}. If larger angular momenta are possible for other theory parameter values, though, this might make the deformed first-order scalarization more astrophysically relevant.}
	\label{fig:M_rho_rot}
\end{figure}

Let us turn now to the effect of uniform rotation when the angular velocity parameter $\Omega \neq 0$ in \eqref{eq:metric}, and again concentrate on the deformed first-order phase transition. To start with, we remind the reader that one can use the turning point  $d\Mb/d\epsc=0$ and the corresponding cusps as an indication for change in the stability of the solutions only along sequences with constant angular momentum $J$. 

The baryon mass as a function of central energy density is plotted in Fig.~\ref{fig:M_rho_rot} for two different sets of theory parameters $(\beta,m_{\varphi})$. The top panel is the same as the $(\beta=-18,m_{\varphi}=6.65 \times 10^{-13} {\rm eV})$ case from the top panel of Fig.~\ref{fig:alpha0_mphi_neq_0} in the previous section, but extended to nonzero $\Omega$. The theory parameters in the bottom panel are chosen such that the bifurcation-like point is at a relatively higher stellar mass. This is mainly because of astrophysical relevance as we will discuss later. Since we keep the angular momentum fixed for all rotating star sequences within each plot for uniform comparison to the non-rotating solutions, the maximum rotation rate we can consider is limited by the Keplerian limit of the lowest mass stars\footnote{The angular momentum at the Kepler limit decreases with decreasing neutron star mass for the parameter range considered here.}. Hence, we had to limit our constant $J$ sequences to a low value of $J=0.025$ in the top panel of Fig.~\ref{fig:M_rho_rot} while moderate $J=0.2$ could be reached in the bottom one.

For the values of $J$ we plot, rotation does not affect the qualitative behavior of the deformed first-order phase transition and does not change how $\alpha$ affects the transition from deformed first-order to deformed second-order in an appreciable way. On the more quantitative side, however, the deformed first-order phase transition disappears for much smaller values of $\alpha$  in the bottom panel of Fig.~\ref{fig:M_rho_rot} compared with the top one. This is related to the fact that the dip in $\Mb(\varepsilon_\textrm{c})$ in the bottom panel is already shallower for $\alpha=0$, which in turn results from the particular choice of theory parameters $(\beta, m_{\varphi})$. 

A main feature in Fig.~\ref{fig:M_rho_rot} and in all the other rotation cases we checked is that rotating configurations have higher stellar masses compared to the static ones for the same central energy density. Furthermore, this increase is larger for larger $J$. Hence, the bifurcation-like point and the local minimum of the $\Mb(\epsc)$ curve (if it exists) are shifted upwards as well, as much as $5\%$ for $J=0.2$ on the bottom panel. 

Recall that the interesting phenomenon of having metastable solutions at the same baryon mass and possible transitions between these occur near the bifurcation point, hence it is important that the stellar mass at this point is astrophysically relevant. In this sense, the increase of the stellar masses due to rotation is desired, since the stars near the bifurcation-like point has masses typically well below the solar mass $M_\odot$. If one picks the values of $(\beta,m_\varphi,\alpha)$ where the bifurcation-like point is already high, then there is a secondary enhancement due to the possibility of having even higher $J$ values, since the mass shedding limit typically increases with compactness of the star, which in turn increases with stellar mass.  Fast rotation might increase the physical relevance of deformed first-order scalarization in this sense, so a more detailed future study of the whole $(\beta,m_\varphi,\alpha, \Omega)$ parameter space might be useful. 

On the other hand, existing results indicate that such expectations are best tempered. \textcite{Unluturk:2025zie} found the bifurcation-like point to be less than roughly $\Mb\sim 0.75 M_\odot$ for any combination of $(\beta,m_\varphi)$, when nonrotating stars with $\alpha=0$ were considered. Even a $20\%$ rotation-induced increase in this mass value, which would be considerably higher than what we observed for any set of theory parameters in our analysis, would still put the bifurcation-like point at a lower mass than what has been astrophysically interesting for neutron stars \cite{Freire:2024adf,Manchester2005ATNF,ATNFOnline}. Note, though, that the recent discovery of a $0.77 M_\odot$ neutron star \cite{Doroshenko2022LightNS} might change the picture if more objects like this are observed in the future.

\begin{figure}[]
    \includegraphics[width=0.45\textwidth]{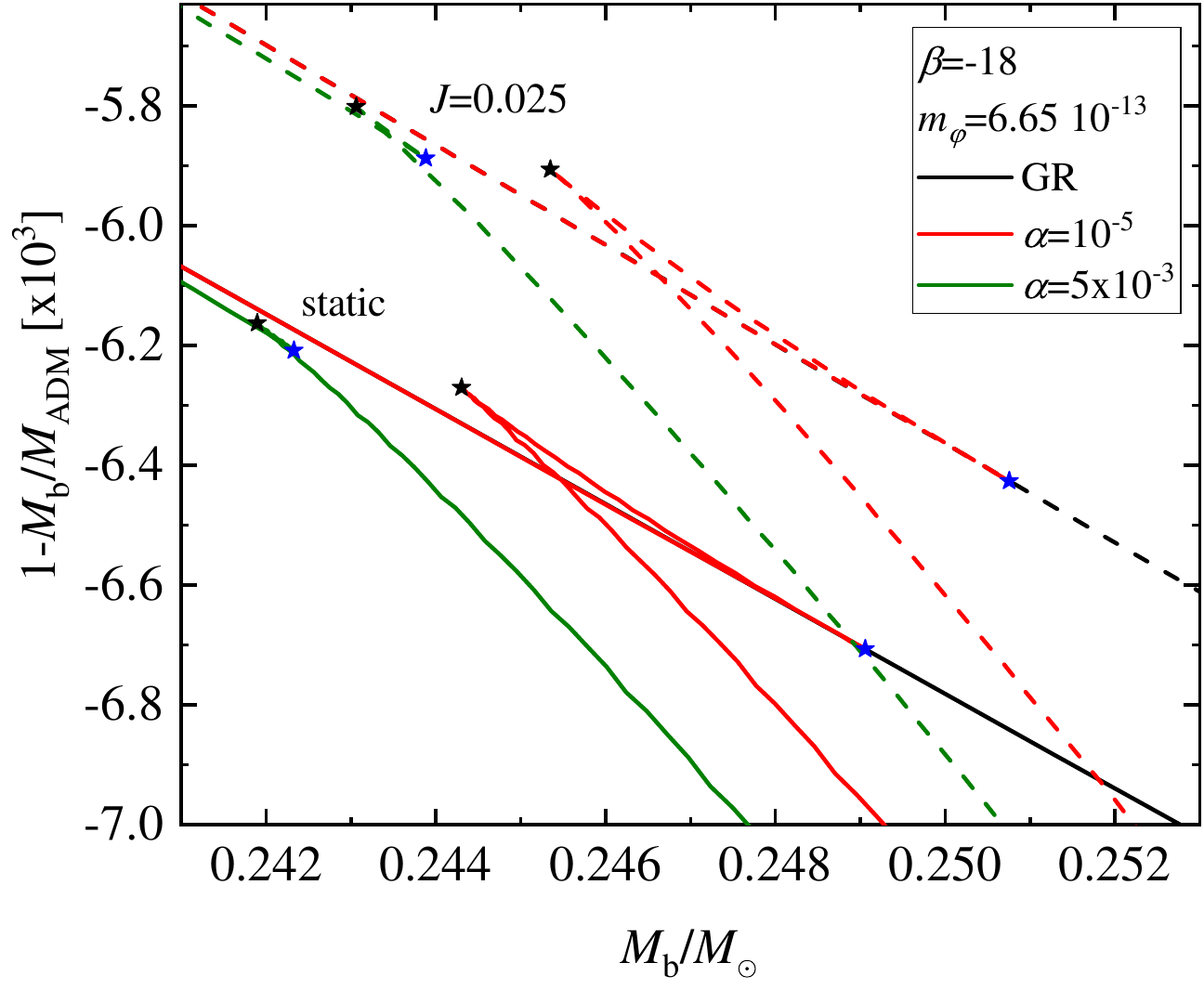}
    \caption{The binding energy as a function of the baryon mass for static and rotating solutions with $\beta =-18$, $m_{\varphi}\simeq 6.65 \times 10^{-13}$ eV and different values for $\alpha$. The stars mark the cusps and they coincide with the ones shown in the top panel of Fig. \ref{fig:M_rho_rot}.}
	\label{fig:M_b_18_Eb}
\end{figure}
Lastly, the behavior of the binding energy and the global stability of the rotating solutions can be found in Fig.~\ref{fig:M_b_18_Eb}, which closely follows the nonrotating case in Fig.~\ref{fig: cusps nonrotating}.

\section{Conclusions}
\label{sec: conclusions}
We extended the research on first-order phase transition in scalar-tensor theories in two main directions: the effect of the linear term $\alpha$ in the conformal coupling and the rotation of the object that scalarizes. We demonstrated that the basic phenomenological view of the scalarization phase transition can be adapted to the presence of a nonzero $\alpha$, and it successfully describes how both the first and the second-order phase transition are deformed from the previously studied examples with $\alpha=0$. In addition, this analysis turned out to be a useful tool for exploring the richness of the solution space. As for rotation, our expectation of an increase in the phase transition stellar masses with the increase of the angular momentum was validated. This is important since the stellar masses where first-order phase transitions are theoretically observed are low compared to most current measurements for neutron star\cite{Unluturk:2025zie}. The increase due to rotation turned out to be moderate, though, and it provided a limited move towards more astrophysically relevant neutron star configurations.

Our analysis based on the Landau ansatz~\eqref{eq: energy landau ansatz} which incorporates the effects of a nonzero $\alpha$ turned out to be an excellent predictor for the phase transition picture on one hand, and a telltale indicator for the richness of the solution space on the other. We could predict up to 5 equilibrium solutions at the same baryon mass, having both positive and negative signs of the scalar field, where two of these solution branches are unstable. The number of solutions are further predicted to change from 5 to 3 and finally to only 1 for low, intermediate, and high values for $\alpha$. This whole description was faithfully replicated by the numerical solutions we found. Though, we should note that whether both the positive and the negative scalar field branches could be realized in astrophysical scenarios remains an open question, and may depend on cosmological boundary conditions in the far region. Also recall that if one relies purely on the outcome of the standard numerical procedures, the full richness of the equilibrium solutions may remain unexplored without the phase transition analysis. Thus, we could recover all the branches thanks to the guidance of the phenomenological picture in Sec.~\ref{sec:phase transition}. Hence, we believe that a basic understanding of the phase transition picture is essential to understand the full spectrum of solutions in any study of scalarization.  

The change in the number of equilibrium solutions with $\alpha$ which we described above also has direct consequences for the phase transition structure of scalarization. Our results show that small values of $\alpha$ preserve the phase transition picture. However, when we further increase $\alpha$, the picture changes qualitatively. For example, what was originally a first-order scalarization with 5 equilibria becomes one with 3 equilibria, closely resembling a slightly deformed version of second-order scalarization for $\alpha=0$. More radically, $\alpha$ values greater than a critical one that depends on the scalar coupling parameter $\beta$ and mass $m_\varphi$ (in absolute value) can lead to a solution space with a single, highly scalarized branch, where there is no phase transition. This is a regime that we can interpret as the $\alpha$ term dominating over $\beta$, hence it is more akin to the BD theory rather than the DEF model where scalarization was originally discovered.

We did not have phenomenological predictions for the effect of neutron star rotation on the phase transition picture, unlike the case of $\alpha$, yet it is well known that both scalarized and unscalarized solutions tend to have higher masses for the same central energy density with increasing angular momentum. Within the rates of uniform rotation we studied, there were no qualitative changes in the phase transition picture. We observed the natural increase in the mass of the solutions associated with  rotation as expected, which made the solutions near the bifurcation-like point more astrophysically relevant, but only slightly so. The more interesting case of deformed first-order scalarization with its 5 equilibrium neutron star configurations at the same baryon mass only occurred at relatively low baryon masses in our search, similarly to what has been observed for $\alpha=0$~\cite{Unluturk:2025zie}. Hence, only relatively low angular momenta $J$ can be supported by such stars before reaching the mass-shedding limit, which in turn leads to only modest increases in the stellar mass due to rotation. Another factor in the prominence of the deformed first-order scalarization is the depth of the bucket in the $\Mb(\epsc)$ curves, but this also did not change substantially with rotation. In short, with our current choice of conformal factor ${\cal A}(\varphi)$ and EOS, rotation does not seem to increase the astrophysical relevance of first-order scalarization in a significant way compared to \textcite{Unluturk:2025zie}. 

So far, we have considered more conventional cases of astrophysical relevance, but let us also mention that a neutron star with a mass as low as $0.77 M_\odot$ has already been discovered \cite{Doroshenko2022LightNS}. Moreover, there are scenarios where neutron stars with masses well below the solar mass have been considered in relation to accretion disks and partial disruption events~\cite{Metzger:2024ujc,Stephens:2011as,East:2011xa}. Overall, there is evidence that even the lower stellar mass first-order scalarization effects which we have studied here have a realistic chance to be tested in astrophysical systems.

We end by reminding the reader that we still had specific assumptions about our STT and the scalarizing object, a neutron star, despite significantly generalizing the state of the art which only considered spherically symmetric configurations with $\alpha =0$. The phenomenological understanding of the scalarization phase transition picture can be widened much further, for example to black hole scalarization in theories where the scalar field couples to curvature terms, for which some initial steps have already been taken~\cite{Herdeiro:2026sur}. Moreover, even though we considered a realistic EOS that is commonly employed for neutron stars, other choices, such as those that predict matter phase transitions near the stellar core, might also lead to a different picture. Whether such considerations can provide further astrophysically interesting systems remains to be seen.

\acknowledgments
We thank O\u{g}uzhan K. Yamak for his comments on the numerical solutions. K.S and S.Y. are supported by the European Union-NextGenerationEU, through the National Recovery and Resilience Plan of the Republic of Bulgaria, project No. BG-RRP-2.004-0008-C01. F.M.R is supported by the Scientific and Technological Research Council of Turkey (T\"UB\.ITAK) Grant
Number 122F097. D.D. acknowledges financial support via an Emmy Noether Research Group funded by the German Research Foundation (DFG) under grant no. DO 1771/1-1, by the Spanish Ministry of Science and Innovation via the Ram\'on y Cajal programme (grant RYC2023-042559-I), funded by MCIN/AEI/ 10.13039/501100011033, and by the Spanish Agencia Estatal de Investigaci\'on (grant PID2024-159689NB-C21) funded by the Ministerio de Ciencia, Innovaci\'on y Universidades. We acknowledge Discoverer PetaSC and EuroHPC JU for awarding this project access to Discoverer supercomputer resources.

\bibliography{bibliography}

\end{document}